\documentclass[12pt, draftcls, letterpaper, onecolumn]{IEEEtran}
\usepackage{cite}

\ifCLASSINFOpdf
   \usepackage[pdftex]{graphicx}
   \DeclareGraphicsExtensions{.pdf,.jpeg,.png}
\else
   \usepackage[dvips]{graphicx}
   \DeclareGraphicsExtensions{.eps}
\fi
\usepackage[cmex10]{amsmath}
\begin{document}
%
\title{Complexity Adjusted Soft-Output Sphere Decoding by Adaptive LLR Clipping}
%
%
%
\author{Konstantinos~Nikitopoulos,~\IEEEmembership{Member,~IEEE,}~and~Gerd~Ascheid,

~\IEEEmembership{Senior Member,~IEEE}
\thanks{The authors are with the Faculty of Electrical Engineering and Information Technology, RWTH Aachen University, Aachen, Germany (e-mail: Konstantinos.Nikitopoulos@iss.rwth-aachen.de).}
\thanks{This work has been supported by the UMIC Research Center, RWTH Aachen University.}}
%
%
\markboth{The final version of this work appears in IEEE Communications Letters}{}
%
\maketitle
\begin{abstract}
  A-posteriori probability (APP) receivers operating over multiple-input, multiple-output channels provide enhanced bit error rate (BER) performance at the cost of increased complexity. However, employing full APP processing over favorable transmission environments, where less efficient approaches may already provide the required performance at a reduced complexity, results in unnecessary processing. For slowly varying channel statistics substantial complexity savings can be achieved by simple adaptive schemes. Such schemes track the BER performance and adjust the complexity of the soft output sphere decoder by adaptively setting the related log-likelihood ratio (LLR) clipping value.
\end{abstract}


%
\IEEEpeerreviewmaketitle

\section{Introduction}
%
%
%
%
\IEEEPARstart {M}{ultiple} antennae systems with spatial multiplexing offer increased spectral efficiency and therefore, they have been adopted by several upcoming wireless communication standards like IEEE 802.11n and IEEE 802.16e. \emph{A-posteriori probability} (APP) receivers, which exploit soft information content, can efficiently decode such schemes but at the cost of increased processing requirements. Accurate (full complexity) APP receiver processing may be demanded for achieving the required bit error rate (BER) performance when transmitting over unfavorable channels (i.e., ill conditioned and of low signal-to-noise ratio (SNR)). However, it may be unnecessary for favorable transmission conditions where the required performance can be achieved even by approximate, reduced soft information processing.


In this context, a low overhead adaptive approach is proposed aiming at avoiding such unnecessary soft-output detection processing. It targets practical scenarios where the channel statistics lead to only slowly varying (over subsequent code blocks) \emph{average} achievable BER performance. The proposed output detector is realized by means of the depth-first sphere decoder (SD) of \cite{ETH_SS} which can ensure the (exact) \emph{max-log} MAP performance, when required. The scheme adjusts the processing requirements of the SD  to the target-error-rate (TER) performance by changing the \emph{log-likelihood ratio} (LLR) clipping value after \emph{evaluating} (i.e., tracking) the provided BER performance. In \cite{ETH_SS,Mylly10} the fundamental idea of tuning the SD complexity and the performance via LLR clipping is demonstrated via extensive simulations. However, no discussion has been made on how to practically set the LLR clipping value in order to adjust on-the-fly the receiver's processing requirements to the TER.

The proposed approach does not require the exact relationship between the LLR clipping value and the resulting performance. Therefore, it is generally applicable to any transmission scenario (i.e., channel code, modulation, SNR, etc.) without necessitating any kind of \emph{performance prediction} (i.e., analytical or by extensive simulations) burden. In this work, the applicability of such adaptive schemes is demonstrated, without claiming any optimality for the specific tracking algorithm.

\section{APP Receiver Processing for MIMO Systems}

In MIMO transmission with  ${M_T}$ transmit and ${{M}_{R}}\ge {{M}_{T}}$ receive antennas, at the $u$-th MIMO channel utilization, the interleaved coded bits are grouped into blocks ${B_{t,u}}$ ($t = 1,...,{M_T}$ and $u = 1,...,{U}$ with ${U}$ being the number of channel utilizations per code block) in order to be mapped onto symbols ${s_{t,u}}$ of a constellation set $S$ of cardinality $\left| S \right|$. The bipolar $k$-th bit resides in block ${B_{{\left\lceil {{k \mathord{\left/
 {\vphantom {k {{{\log }_2}\left| S \right|}}} \right.\kern-\nulldelimiterspace} {{{\log }_2}\left| S \right|}}} \right\rceil },u}}$ and the blocks ${B_{t,u}}$ are mapped onto the symbols ${s_{t,u}}$  by a given mapping function (e.g., Gray mapping). The corresponding received ${M_R}\times{1}$ vector ${\bf{y}}_{u}$ is, then, given by
\begin{equation}
{\bf{y}}_{u} = {\bf{H}}_{u}{\bf{s}}_{u} + {\bf{n}}_{u},	
\end{equation}
where ${\bf{H}}_{u}$ is the ${M_R}\times{M_T}$ complex channel matrix which can be QR decomposed into ${\bf{H}}_{u} = {\bf{Q}}_{u}{\bf{R}}_{u}$, with ${\bf{Q}}_{u}$ a unitary \mbox{${M_R}\times{M_T}$} matrix and ${\bf{R}}_{u}$ an ${M_T}\times{M_T}$ upper triangular matrix with elements ${R_{i,j,u}}$ and real-valued positive diagonal entries. By ${\bf{s}}_{u} = {\left[ {{s_{1,u}},{s_{2,u}},...,{s_{{M_T,u}}}} \right]^T}$ the transmitted symbol vector is denoted, while ${\bf{n}}_{u}$
  is the noise vector consisting of i.i.d., zero-mean, complex, Gaussian  samples with variance $2\sigma _n^2$.

  The soft-output detector calculates the \emph{a-posteriori} LLRs for all the symbols residing in the frame to be decoded. Assuming statistically independent bits (due to interleaving) and under the \emph{max-log} approximation, due to the QR decomposition the LLR calculation problem can be reformulated to the following constraint tree-search problem known as SD \cite{Hochwald03}
 \[{{L}_{D}}\left( {{c}_{b,i,u}} \right)\approx \frac{1}{2\sigma _{n}^{2}}\underset{{{\mathbf{s}}_{u}}\in S_{b,i,u}^{-1}}{\mathop{\min }}\,{{\left\| \mathbf{y}{{'}_{u}}-{{\mathbf{R}}_{u}}{{\mathbf{s}}_{u}} \right\|}^{2}}-\]
  \begin{equation}
\frac{1}{2\sigma _{n}^{2}}\underset{{{\mathbf{s}}_{u}}\in S_{b,i,u}^{+1}}{\mathop{\min }}\,{{\left\| \mathbf{y}{{'}_{u}}-{{\mathbf{R}}_{u}}{{\mathbf{s}}_{u}} \right\|}^{2}}
 \end{equation}
 where  $c_{b,i,u}$ is the $b$-th bit of the $i$-th entry of ${\bf{s}}_{u}$, $S_{b,i,u}^{ \pm 1}$ are the sub-sets of possible ${{\bf{s}}_u}$ symbol sequences having the $b$-th bit value of their $i$-th ${\bf{s}}_{u}$ entry equal to $ \pm 1$ and ${\bf{y'}}_{u} = {{{\bf{Q}}_{u}}^H}{\bf{y}}_{u} = {\left[ {y{'_{1,u}},y{'_{2,u}},...,y{'_{{{M_T,u}}}}} \right]^T}$. Then the corresponding tree search problem can be solved according to \cite{ETH_SS}.

The resulted soft information is de-interleaved and fed to the soft-input, soft-output (SISO) channel decoder as \emph{a-priori} information, with ${{\tilde{L}}_{A}}\left( {{{\tilde{c}}}_{k}} \right)$ denoting the \emph{a-priori} information of the ${\tilde{c}}_{k}$ encoded bit. This information is employed for calculating the channel decoder's \emph{a-posteriori} soft information ${{\tilde{L}}_{D}}\left( {{{\tilde{c}}}_{k}} \right)$ whose sign provides the corresponding decoded bit ${{\hat{c}}_{k}}$.
Practically, this \emph{a-posteriori} information can be efficiently calculated by the BCJR-MAP algorithm \cite{BCJR}.

By employing the \emph{max-log} approximation (for easier inspection and without loss of generality), and since
\begin{equation}
P\left[ {{{\tilde{c}}}_{i}} \right]=\frac{\exp \left( -{\left| {{{\tilde{L}}}_{A}}\left( {{{\tilde{c}}}_{i}} \right) \right|}/{2}\; \right)}{1+\exp \left( -\left| {{{\tilde{L}}}_{A}}\left( {{{\tilde{c}}}_{i}} \right) \right| \right)}\exp \left( {{{{\tilde{c}}}_{i}}{{{\tilde{L}}}_{A}}\left( {{{\tilde{c}}}_{i}} \right)}/{2}\; \right),
\end{equation}
after extracting some mutually exclusive terms, it can be easily shown that
\[{{\tilde{L}}_{D}}\left( {{{\tilde{c}}}_{k}} \right)\approx \frac{1}{2}\underset{\mathbf{\tilde{c}}:\tilde{C}_{k}^{+1}}{\mathop{\max }}\,\left\{ \sum\limits_{i=1}^{k}{\left( {{{\tilde{c}}}_{i}}{{{\tilde{L}}}_{A}}\left( {{{\tilde{c}}}_{i}} \right)-\left| {{{\tilde{L}}}_{A}}\left( {{{\tilde{c}}}_{i}} \right) \right| \right)} \right\}-\]
\begin{equation}
\frac{1}{2}\underset{\mathbf{\tilde{c}}:\tilde{C}_{k}^{-1}}{\mathop{\max }}\,\left\{ \sum\limits_{i=1}^{k}{\left( {{{\tilde{c}}}_{i}}{{{\tilde{L}}}_{A}}\left( {{{\tilde{c}}}_{i}} \right)-\left| {{{\tilde{L}}}_{A}}\left( {{{\tilde{c}}}_{i}} \right) \right| \right)} \right\},
\end{equation}
with $\tilde{C}_{k}^{\pm 1}$ being the set of possible encoded sequences $\mathbf{\tilde{c}}$, of length $K$, with their $k$-th bit equal to $\pm 1$. The candidate sequences for maximizing each of the terms in (4) are those with the corresponding non-positive sums being as close to zero as possible. Therefore, it is not expected that highly unlikely bits (i.e., with high $\left| {{{\tilde{L}}}_{A}}\left( {{{\tilde{c}}}_{i}} \right) \right|$ value and sign opposite to ${{{\tilde{L}}}_{A}}\left( {{{\tilde{c}}}_{i}}\right)$, see (3)) will belong to those sequences (except possibly the bit under constraint $\tilde{c}_{k}$) since they contribute with highly negative values. In addition, since the candidate bits (except for $\tilde{c}_{k}$) are expected to contribute with values close to zero, bounding the $\left| {{{\tilde{L}}}_{A}}\left( {{{\tilde{c}}}_{k}} \right) \right|$ value of the bit under constraint by a large value will not affect the final sign of (4) and, consequently, the decoded bit. Therefore, approximate calculation of the \emph{a-priori} information of the the highly unlikely bits is not expected to significantly affect the resulting BER. In the same way, highly likely bits (i.e., with high $\left| {{{\tilde{L}}}_{A}}\left( {{{\tilde{c}}}_{i}} \right) \right|$ value and of the same sign with ${{{\tilde{L}}}_{A}}\left( {{{\tilde{c}}}_{i}}\right)$) will contribute with a zero value independently of their actual \emph{a-priori} information (see (4)). The last observations lead to the conclusion that approximate (and thus of lower complexity) calculation of the strong soft information, (i.e., of high $|{{\tilde{L}}_{A}}\left( {{\tilde{c}}_{k}} \right)|$) is not expected to significantly affect the outcome of the SISO channel decoder and, therefore, the provided BER.

\section{Adaptive LLR Clipping}
The problem to be addressed is to adaptively find the minimum clipping value, and thus to minimize the SD complexity, which does not jeopardize the TER performance. To this direction, a coarse LLR clipping value $L_{cl}^{(0)}$ is initially set, large enough to safely preserve the TER. Then, $L_{cl}$ is adaptively reduced so that the provided performance reaches the TER. In detail, the following steps are performed:
\begin{enumerate}
\item
\emph{BER Evaluation}:
According to \cite{Land00} the error probability of the hard decision of the bit $c$ with \emph{a-posteriori} LLR $L_{c}$ is
 \begin{equation}
{{P}_{e}}(c)={{\left( 1+\exp \left( \left| L\left( c \right) \right| \right) \right)}^{-1}}.
 \end{equation}
Since the average error rate is dominated by the bits with the smaller LLR magnitudes, the BER of the $m$-th code block can be approximated as
 \begin{equation}
\hat{P}_{b}^{(m)}\approx \frac{1}{{{N}}}\sum\limits_{r=1}^{{{n}}}{{{P}_{e}}\left( \overset{\lower0.5em\hbox{$\smash{\scriptscriptstyle\smile}$}}{c}_{r,m}^{(I)}\right),}\begin{matrix}{}&{{n}}\le\\\end{matrix}{{N}}
  \end{equation}
 with $\overset{\lower0.5em\hbox{$\smash{\scriptscriptstyle\smile}$}}{c}_{r,m}^{(I)}$ being the information bit of the $m$-th code block with the $r$-th smallest $\left| {{{\tilde{L}}}_{D}}\left( {{{\tilde{c}}}_{k}} \right) \right|$ value, $N$ being the number of information bits and with the accuracy of the estimate increasing with $n$.
\item
\emph{Clipping Value Initialization}:
For the bits which meet the TER constraint already before channel decoding (i.e., $\left| {{{\tilde{L}}}_{A}}\left( \tilde{c}_{k} \right) \right| > {{\tilde{L}}_{TER}}=\ln (TE{{R}^{-1}}-1)\approx -\ln (TER)$, see (5)) their average BER performance after channel decoding is expected to be even better. Therefore, the $L_{cl}^{(0)}$ value can be set equal to ${\tilde{{L}}_{TER}}$. This initialization does not compromise the TER performance since, as discussed, high magnitude LLR clipping does not significantly affect the SISO decoding. This is verified by simulations in Section IV.
\item
\emph{Clipping Value Adaptation}:
The following simple adaptive algorithm can be used to adjust the LLR clipping value to the TER performance
 \begin{equation}
L_{cl,c}^{(m)}=L_{cl}^{(m-1)}-\mu \left[ \ln \left( TER \right)-\ln \left( \hat{P}_{b}^{(m-1)} \right) \right]
 \end{equation}
  \begin{equation}
L_{cl}^{(m)}=\max \left\{ \min \left\{ {{L}_{TER}},L_{cl,c}^{(m)} \right\},{{\left| L \right|}_{\min }} \right\}
 \end{equation}
with $\mu$ being the step size parameter and ${{\left| L \right|}_{\min }}$ being the minimum possible LLR magnitude determined by the fixed point accuracy. Then, by (8) non-positive $L_{cl}$ values are avoided ($\max$ operation). Additionally, unnecessary $L_{cl}$ (and thus complexity) increase is avoided for those cases where the best achievable error rate performance is worse than the TER ($\min$ operation).
\end{enumerate}
     If the $\exp(|L(c)|)$ and the $\ln(\hat{P}_{b})$ operations of (5) and (7) are implemented by means of look-up-tables (which can be efficiently done for the $L(c)$ and $\hat{P}_{b}$ ranges of interest), the main processing overhead of the proposed approach can be approximated to be $n$ (Eq. 5) + $1$ (Eq. 6) + $1$ (Eq. 7)  = $n+2$ real divisions per code block. However, as shown in Section IV, a small $n$ ($\le 50$) provides adequate accuracy. Therefore, the induced overhead is very small compared to the (per code block) SD complexity gain of Section IV.

\section{Simulations}
A $4\times4$ MIMO system is assumed, operating over a spatially and temporally uncorrelated Rayleigh flat-fading channel. The encoded bits are mapped onto 16-QAM via Gray coding. A systematic  ${(5/7)_8}$
 recursive convolutional code of rate 1/2 is employed with a code block of $N=1152$ information bits. The log-MAP BCJR algorithm has been employed for SISO channel decoding. Tracking takes place over 100 consequent frames and $\mu=0.1$ unless otherwise specified.

 In Fig. 1 the BER performance of the proposed approach is shown for various TER values ($10^{-4}$, $10^{-3}$, $10^{-2}$) and $n=50$ and $N$. For $n=50$ the estimated BER is slightly lower (better) than for $n=N$ (see (6)). For high TER values this estimation difference can lead to significantly larger tracking steps (see (7)) and therefore to faster convergence. Then, significant average performance difference can be depicted as in Fig. 1. To validate the efficiency of the $L_{cl}$ initialization, $\mu=0$ is also considered. Then, as expected, a performance divergence is observed only when the TER performance is worse (higher) than the best achievable. However, the resulted degraded performance still significantly outperforms the TER which denotes that a great amount of unnecessary processing takes place. The last is reduced for $\mu=0.1$ and $n=N$, but with a provided BER which can be slightly worse than the TER in the high SNR regime and for high TER values (see TER=$10^{-4}$ and SNR $>12$ dB) due to the oscillations of the clipping value around the optimal. In practice, this performance loss can be reduced by setting a smaller $\mu$ value, at the cost of reduced convergence speed, or by setting a TER value slightly better than the actual.

 In Fig. 2 the complexity is depicted in terms of average number of visited nodes per MIMO channel utilization (i.e., 144 MIMO channel utilizations form a code block). Significant complexity gains can be observed over the whole SNR range even for $\mu=0$ when compared to the non-clipped case, as in \cite{ETH_SS}, despite the absence of the exact relationship between the LLR clipping and the resulting performance. For example, for SNR=14 dB and TER=$10^{-4}$ there is gain of 92\% compared to the non-clipped SD (requiring $1.49\cdot 10^3$ visited nodes, by simulations not depicted here). Substantial complexity gains can be further achieved by the proposed scheme when the best achievable performance is significantly better than the TER. For example, as shown in Fig. 2, an additional gain of about 53\% is provided for SNR=14 dB, TER=$10^{-4}$ and $\mu=0.1$, while the additional gain is about 40\% for TER=$10^{-2}$.

 As discussed in Section I, the proposed approach avoids any performance prediction burden. However, it would be interesting to examine the additional complexity gains under optimal $L_{cl}$ knowledge, even without considering the computational overhead to calculate it. Fig. 2 shows that a clipping value of $\ln \left( {1}/{{{10}^{-2}}-1}\; \right)$ without tracking achieves a TER of $10^{-3}$ at 12 dB. Then, additional complexity gains of only 5\% could be achieved compared to the proposed scheme targeting the same TER (see Fig. 2). The same clipping value achieves a TER of $10^{-4}$ at 13.75 dB. In this case, the proposed approach would be even slightly less complex (due the small BER degradation) and with a gain of about 50\% compared to the $\mu=0$, TER=$10^{-4}$ case.

\section{Conclusion}
It is demonstrated that the complexity of the soft-output sphere decoding can be efficiently adjusted to the required BER performance by simple schemes which track the provided performance and adaptively adjust the LLR clipping value, when the channel statistics vary slowly. Then, substantial complexity savings can be achieved without requiring the exact relationship between the clipping value and the resulting BER performance.


%




\ifCLASSOPTIONcaptionsoff
  \newpage
\fi



%


\bibliographystyle{IEEEtran}
\bibliography{JustNeededSD_Bib}

\begin{thebibliography}{1}
\providecommand{\url}[1]{#1}
\csname url@samestyle\endcsname
\providecommand{\newblock}{\relax}
\providecommand{\bibinfo}[2]{#2}
\providecommand{\BIBentrySTDinterwordspacing}{\spaceskip=0pt\relax}
\providecommand{\BIBentryALTinterwordstretchfactor}{4}
\providecommand{\BIBentryALTinterwordspacing}{\spaceskip=\fontdimen2\font plus
\BIBentryALTinterwordstretchfactor\fontdimen3\font minus
  \fontdimen4\font\relax}
\providecommand{\BIBforeignlanguage}[2]{{%
\expandafter\ifx\csname l@#1\endcsname\relax
\typeout{** WARNING: IEEEtran.bst: No hyphenation pattern has been}%
\typeout{** loaded for the language `#1'. Using the pattern for}%
\typeout{** the default language instead.}%
\else
\language=\csname l@#1\endcsname
\fi
#2}}
\providecommand{\BIBdecl}{\relax}
\BIBdecl

\bibitem{ETH_SS}
{C. Studer, A. Burg, and H. B\"olcskei}, ``{Soft-output sphere decoding:
  algorithms and VLSI implementation},'' \emph{{IEEE J. Sel. Areas Commun.}},
  vol.~26, pp. 290--300, 2008.

\bibitem{Mylly10}
{M. Myllyll\"a, M. Juntti, and J. Cavallaro}, ``{Implementation aspects of list
  sphere detector algorithms for MIMO-OFDM systems},'' \emph{{Signal
  Processing, Elsevier}}, vol.~90, no.~10, pp. 2863--2876, Oct. 2010.

\bibitem{Hochwald03}
{B. M. Hochwald, and S. ten Brink}, ``{Achieving near-capacity on a
  multiple-antenna channel},'' \emph{{IEEE Trans. Commun.}}, vol.~51, no.~3,
  pp. 389--399, Mar. 2003.

\bibitem{BCJR}
{L. Bahl, J. Cocke, F. Jelinek, and J. Raviv}, ``{Optimal decoding of linear
  codes for minimizing symbol error rate},'' \emph{{IEEE Trans. Inf. Theory}},
  Mar. 1974.

\bibitem{Land00}
{I. Land, and P. A. Hoeher}, ``{Log-likelihood values and Monte Carlo
  simulation - some fundamental results},'' in \emph{{Proc. of Int. Symposium
  on Turbo Codes and Related Topics}}, 2000.

\end{thebibliography}

  \begin{figure}
  \centering
  \includegraphics[width=1.05\linewidth]{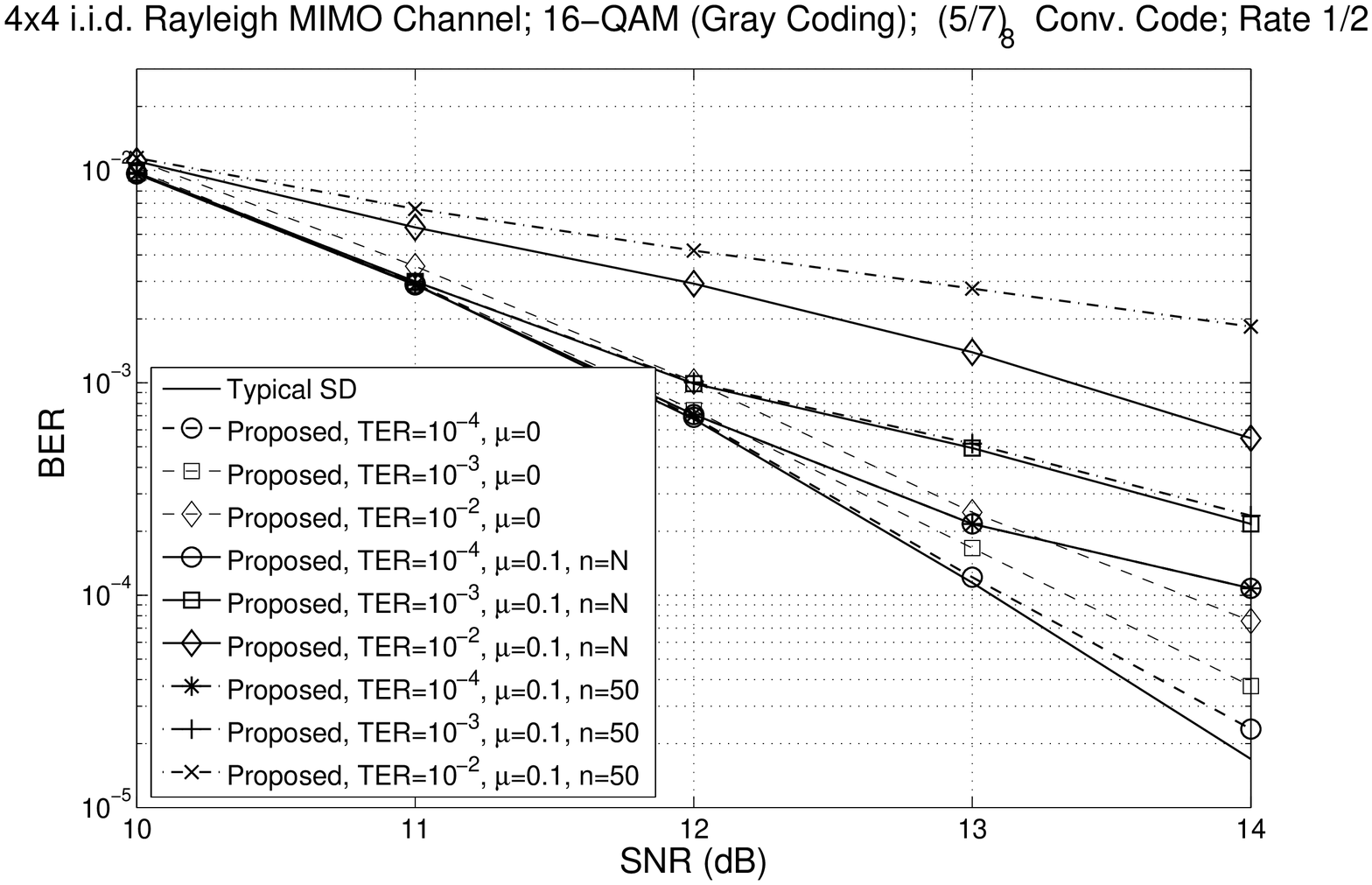}\\
  \caption{BER performance for several TER values.}
\end{figure}

\begin{figure}
  \centering
  \includegraphics[width=1.05\linewidth]{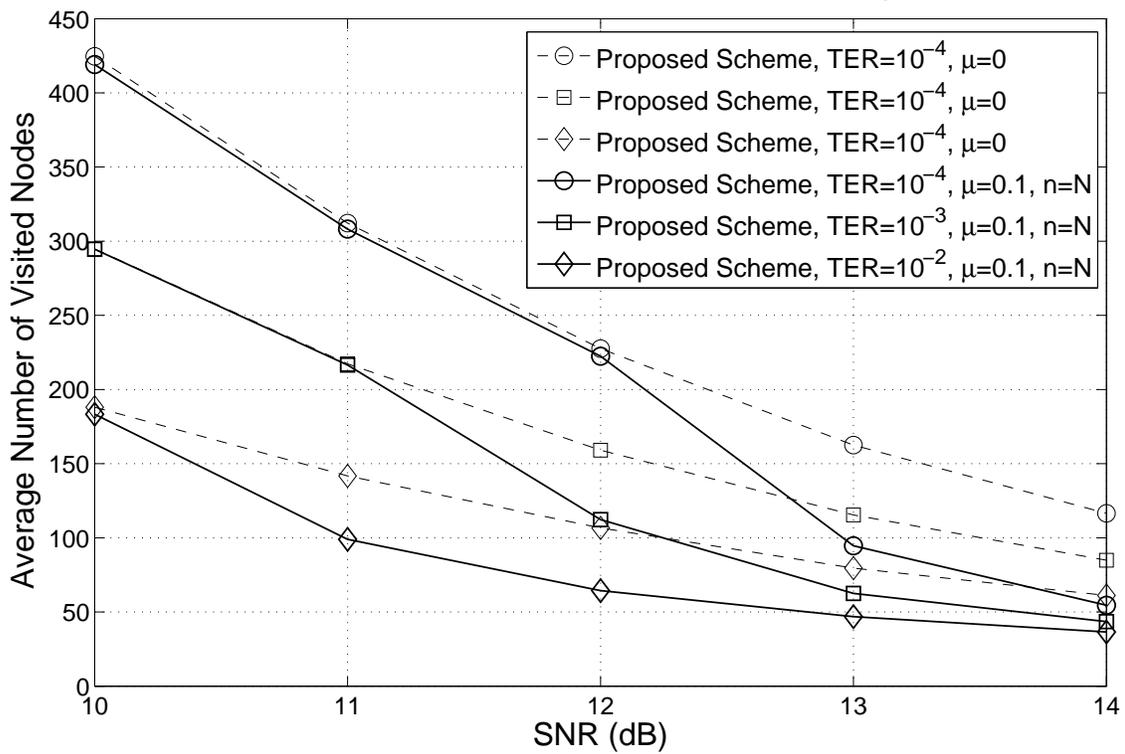}\\
  \caption{Complexity, in terms of average number visited nodes for several TER values.}
\end{figure}


%








\end{document}